\documentclass[]{aa}
\usepackage{graphicx}
\usepackage{txfonts}
\usepackage{psfig}

\begin{document}

\title{Mid-infrared spectral evidence for a luminous dust
       enshrouded source in Arp\,220
\thanks{Based on observations with ISO, an ESA project with 
       instruments funded by ESA Member States (especially the PI 
       countries: France, Germany, the Netherlands and the United 
       Kingdom) and with the participation of ISAS and NASA}}

\author{H.W.W. Spoon\inst{1}
   \and
        A.F.M. Moorwood\inst{2}
   \and
        D. Lutz\inst{3}
   \and
        A.G.G.M. Tielens\inst{1,4} 
   \and
        R. Siebenmorgen\inst{2}
   \and
        J.V. Keane\inst{5}}

\offprints{H.W.W. Spoon ({\tt spoon@isc.astro.cornell.edu})}
   
\institute{Kapteyn Institute, P.O. Box 800, NL-9700 AV Groningen, 
           the Netherlands
       \and
           European Southern Observatory, Karl-Schwarzschild Strasse 2,
           D-85748 Garching, Germany
       \and
           Max-Planck-Institut f\"ur Extraterrestrische Physik (MPE),
           Postfach 1312, D-85741 Garching, Germany
       \and
           SRON, P.O. Box 800, NL-9700 AV Groningen, the Netherlands
       \and
           NASA-Ames Research Center, Mail Stop 245-6, Moffett Field, 
           CA 94035, USA}

\date{Received date; accepted date}

\abstract{We have re-analyzed the 6--12\,$\mu$m ISO spectrum of the 
ultra-luminous infrared galaxy Arp\,220 with the conclusion that it is 
not consistent with that of a scaled up version of a typical 
starburst. Instead, both template fitting with spectra of the 
galaxies NGC\,4418 and M\,83 and with dust models suggest that it is 
best represented by combinations of a typical starburst component, 
exhibiting PAH emission features, and a heavily absorbed dust 
continuum which contributes $\sim$40\% of the 6--12\,$\mu$m flux
and likely dominates the luminosity. Of particular 
significance relative to previous studies of Arp\,220 is the fact that 
the emission feature at 7.7\,$\mu$m comprises both PAH emission and 
a broader component resulting from ice and silicate absorption 
against a heavily absorbed continuum. Extinction to the PAH 
emitting source, however, appears to be relatively low. We 
tentatively associate the PAH emitting and heavily dust/ice absorbed 
components with the diffuse emission region and the two compact 
nuclei respectively identified by Soifer et al. (\cite{Soifer02}) in 
their higher spatial resolution 10\,$\mu$m study. Both the similarity of 
the absorbed continuum with that of the embedded Galactic protostars and 
results of the dust models imply that the embedded source(s) in Arp\,220 
could be powered by, albeit extremely dense, starburst activity. 
Due to the high extinction, it is not possible with the available data 
to exclude that AGN(s) also contribute some or all of the observed 
luminosity. In this case, however, the upper limit measured for its 
hard X-ray emission would require Arp\,220 to be the most highly 
obscured AGN known.

}

\maketitle

\section{Introduction}

The galaxy Arp\,220 (IC\,4553; cz=5450\,km/s; D=73\,Mpc) was originally 
classified by Arp (\cite{Arp66}) as a ``galaxy with adjacent loops''. 
Its optical image (1$\arcsec$ = 352\,pc) shows faint structures, 
reminiscent of tails or loops, suggesting it to be the remnant of a recent 
galaxy merger (Toomre \& Toomre \cite{Toomre72}).
IRAS (1983) increased interest in Arp\,220 through the discovery of its
far-IR luminosity and infrared-to-blue ratio which characterized it as an
extreme member of the ``unidentified infrared sources'' discovered during 
the mission (Houck et al. \cite{Houck84}; Soifer et al. \cite{Soifer84}). 
When later the spectroscopic redshifts 
of these ``unidentified infrared sources'' became available, Arp\,220
turned out to have a similarly large infrared luminosity 
(1.35\,$\times$\,10$^{12}$\,L$_{\odot}$), 
making it the nearest member (by a factor of $\sim$2) of the new class 
of UltraLuminous InfraRed Galaxies (ULIRGs; Sanders et al. \cite{Sanders88}), 
with L$_{\rm IR}\geq$10$^{12}$\,L$_{\odot}$. Numerous studies across all 
wavebands have since examined Arp\,220 in close detail, also showing this
nearest ULIRG to be unusual in some aspects rather than being typical for
the class.

\begin{figure*}[]
 \begin{center}
  \psfig{figure=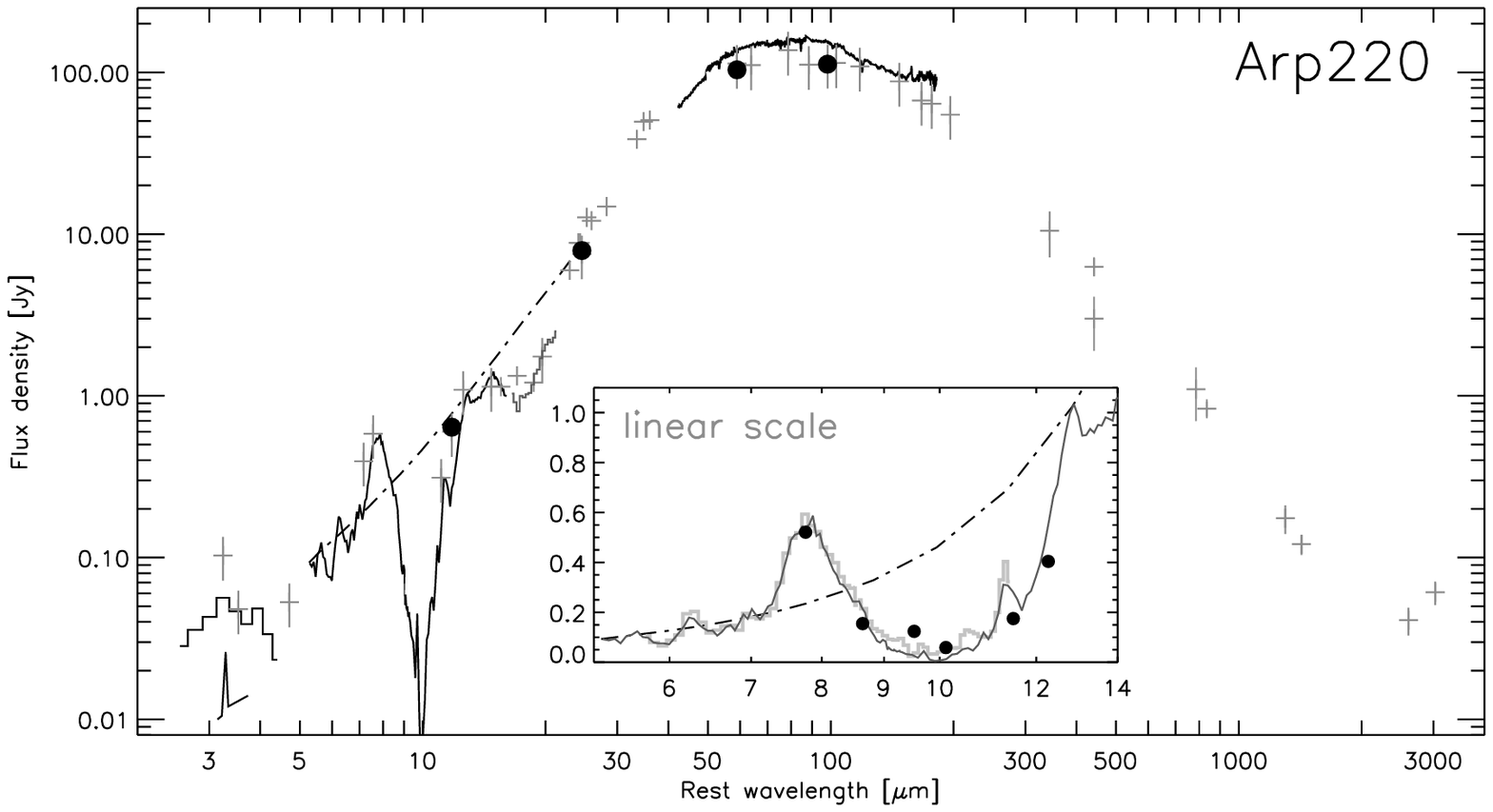,width=17.9cm,angle=0}
 \end{center}
\vspace*{-5mm}
\caption{The 2--3000\,$\mu$m spectrum of Arp\,220. {\it Filled circles}
represent the IRAS fluxes. Spectra shown are: the smoothed 2.4--4.9\,$\mu$m 
ISO-PHT-S spectrum (Spoon et al. \cite{Spoon02}); the 3.2--3.8\,$\mu$m 
CGS4 spectrum obtained in 1.2$\arcsec$ slit 
(Imanishi \& Dudley \cite{Imanishi00}); the 5.0--16\,$\mu$m ISO-CAM-CVF 
spectrum (Tran et al. \cite{Tran01}); the 17--22\,$\mu$m UCL spectrum 
(Smith et al. \cite{Smith89}); the smoothed 45--200\,$\mu$m ISO-LWS 
spectrum (Fischer et al. \cite{Fischer97}).
Other spectral data included in the plot are ISO-PHT photometry
(Klaas et al. \cite{Klaas97}; M.Haas, priv. comm.); ISO-SWS background
subtracted continuum measurements (E.Sturm, priv. comm.); UKIRT and
SCUBA far-IR photometry (Eales et al. \cite{Eales89}; Dunne et
al. \cite{Dunne00},\,\cite{Dunne01}) and mm-observations 
(Anantharamaiah et al. \cite{Anantharamaiah00}). 
The {\it dash-dotted line} is one choice for 
the local continuum in the 5--25\,$\mu$m region (see also 
Fig.\,\ref{mirseds}). The inset shows a comparison of the 
5.8--11.7\,$\mu$m ISO-PHT-S (Spoon et al. \cite{Spoon02}) and the
5.0--16\,$\mu$m ISO-CAM-CVF spectra (shown as {\it grey} and {\it black lines} 
respectively) with the Keck-MIRLIN photometry ({\it filled circles}) of 
Soifer et al. (\cite{Soifer99}).}
\label{arp220_sed}
\end{figure*}

Like most other ULIRGs, Arp\,220 is the product of the interaction of 
two gas-rich disk galaxies (Sanders \& Mirabel \cite{Sanders96}). 
Groundbased observations at 10--30\,$\mu$m suggest that its luminosity
arises in the innermost 250\,pc (Wynn-Williams \& Becklin 
\cite{Wynn-Williams93}).
Radio and mm observations reveal its two nuclei to be surrounded by 
molecular disks of r$\sim$100\,pc, which counterrotate with respect to
each other (Sakamoto et al. \cite{Sakamoto99}). The eastern nucleus
seems to be embedded within an outer gas disk of r$\sim$1\,kpc, which
rotates in the same sense. The western nucleus is connected to the 
eastern nucleus by a thin gas bridge, traced in $\ion{H}{i}$ absorption, 
and appears to lie above the outer gas disk (Mundell et al. \cite{Mundell01}). 
The projected separation of the two nuclei amounts to 345\,pc 
(0.98$\arcsec$; Baan \& Haschick \cite{Baan95}).
High sensitivity VLBI observations disclose the 
presence of multiple compact radio sources dispersed over the two nuclei. 
The knots are consistent with free-free emission from luminous radio 
supernovae expanding in a dense medium (Smith et al. \cite{Smith98,Smith99}).

At shorter wavelengths (in the UV, optical and near-IR) the view towards 
the nuclear components is greatly impaired by strong dust extinction 
of at least A$_{\rm V}$=30--50 (Sturm et al. \cite{Sturm96}). In the 
mid-IR, the dust opacity (A$_{\lambda}$) is a factor of 10--100 less than 
at optical wavelengths and
Smith et al. (\cite{Smith89}) used this property to study the 
nature of the central power source in Arp\,220 in the 8--13\,$\mu$m 
(N-band) and 17--22\,$\mu$m (Q-band) atmospheric windows. 
Based on the weakness of the 11.2\,$\mu$m 
PAH emission band within the deep 9.7\,$\mu$m silicate absorption feature, 
they concluded that only 2--10\% of the total infrared luminosity is
powered by starburst activity, with an obscured AGN responsible for the rest.
Further analysis of the 9.7\,$\mu$m silicate absorption feature led 
Dudley \& Wynn-Williams (\cite{Dudley97}) to conclude, however, that the
obscured power source resembles a scaled-up embedded protostar. 

Not limited to the mid-IR atmospheric windows, ISO spectroscopy revealed 
two pronounced spectral features in the previously unstudied 5--8\,$\mu$m 
range. 
In line with ISO observations of Galactic star forming regions, the two 
features were identified as the 6.2\,$\mu$m and 7.7\,$\mu$m PAH emission 
bands (Genzel et al \cite{Genzel98}). Using the ratio of 7.7\,$\mu$m PAH 
emission to the underlying 7.7\,$\mu$m continuum as a criterium to 
discern starburst- and AGN-dominated galaxies, Genzel et al. 
(\cite{Genzel98}), Lutz et al. (\cite{Lutz98}), 
Spoon et al. (\cite{Spoon98}), Rigopoulou et al. (\cite{Rigopoulou99}) and 
Tran et al. (\cite{Tran01}) classified Arp\,220 as starburst-dominated. 
High angular resolution groundbased N-band spectroscopy has since shown the 
11.2\,$\mu$m PAH emission in the nuclear region to be diffusely distributed 
over the central $\sim$2$\arcsec$ and the starburst associated with the PAH
emission not to be able to account for more than 10--50\% of the bolometric 
luminosity (Soifer et al. \cite{Soifer02}). 
In summary, the mid-IR low-resolution spectral diagnostics appear 
mostly starburst-like but star formation appears quantitatively insufficient 
to account 
for the bolometric luminosity, unless strongly obscured or otherwise
modified. The same is, to a lesser degree, true for the more direct tracing 
of starburst activity by mid-IR fine-structure lines. The ratio of 
$\sim$1/37 of starburst ionizing luminosity and bolometric luminosity, 
derived by Genzel et al. (\cite{Genzel98}) for this source, is about a 
factor 2 less than in comparison starbursts.

Since Arp\,220 is often regarded as a nearby template for dusty galaxies
at high redshift undergoing vigorous star formation (e.g. faint SCUBA
sources), it is imperative to clearly identify its power source(s). 
Despite the quantitative problems with the starburst origin for the 
luminosity, alluded to above, the general consensus since ISO has been 
massive young stars (Genzel \& Cesarsky \cite{Genzel00}). 
However, the infrared luminous 
galaxy NGC\,4945, which shows no outward evidence for an active 
galactic nucleus even in ISO observations (Genzel et al. \cite{Genzel98}; 
Spoon et al. \cite{Spoon00}), has turned out to contain a heavily obscured 
AGN visible only in hard X-rays (Iwasawa et al. \cite{Iwasawa93}; 
Done et al. \cite{Done96}; Guainazzi et al. \cite{Guainazzi00}).
For Arp\,220, BeppoSAX and Chandra observations do not detect a 
similar hard X-ray source (Iwasawa et al. \cite{Iwasawa01}; 
Clements et al. \cite{Clements02}). The only possibility for an 
energetically significant AGN to exist in Arp\,220 would therefore be in the 
form of a deeply embedded source, hidden behind a `Compton-thick' shell of 
N$_{\rm H}\geq$10$^{25}$\,cm$^{-2}$ with a covering factor close to unity
(Iwasawa et al. \cite{Iwasawa01}). The presence of huge amounts of molecular 
gas in the central parts ($\sim$10$^{10}$\,M$_{\odot}$; 
Scoville et al. \cite{Scoville97}; Sakamoto et al. \cite{Sakamoto99}) 
indicates that sufficient obscuring material is indeed 
at hand. And the very large 850\,$\mu$m dust-continuum flux to 7.7\,$\mu$m 
PAH flux (Haas et al. \cite{Haas01}) could mean that the luminosity of this
embedded source is redistributed into the far-IR.

\section{The infrared spectrum of Arp\,220}

The 2--3000\,$\mu$m spectral energy distribution of Arp\,220
(Fig.\,\ref{arp220_sed}) is characterised by a steeply rising
dust continuum, setting in around 4--5\,$\mu$m, and leading up 
to a remarkably strong far-IR flux peak at 60--100\,$\mu$m. 
The most prominent features in the infrared spectrum are the 
silicate absorption features at 9.7\,$\mu$m and 18\,$\mu$m and 
a broad flux peak at 7.8\,$\mu$m.

\subsection{The mid-IR spectrum of Arp\,220}

\begin{figure}[]
 \begin{center}
  \psfig{figure=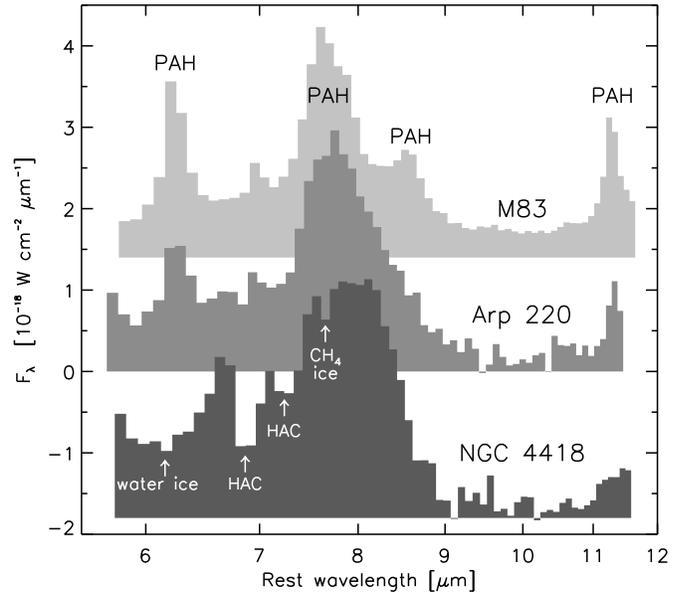,angle=0}
 \end{center}
\vspace*{-5mm}
\caption{A comparison of the ISO--PHT--S spectra of M\,83, Arp\,220 
and NGC\,4418. While the spectrum of M\,83 is dominated by PAH emission 
bands, the spectrum of NGC\,4418 is dominated by absorption bands of
ices and silicates. The spectrum of Arp\,220 shows characteristics of
both. The spectra of M\,83 and NGC\,4418 have been scaled and 
offset.}
\label{photseds}
\end{figure}

In Fig.\,\ref{photseds} we compare the mid-IR spectra of Arp\,220,
NGC\,4418 and the central region of the starburst galaxy M\,83. At first 
sight the three spectra look quite similar. The spectral structure
in the spectrum of NGC\,4418 is, however, the product of strong ice 
and silicate absorptions (Spoon et al. \cite{Spoon01}), whereas the 
spectrum of M\,83 is dominated by the commonly observed emission bands 
of Polycyclic Aromatic Hydrocarbons (PAHs). 
The spectrum of Arp\,220 shows characteristics of both: PAH emission is 
readily detected at 6.2\,$\mu$m and 11.2\,$\mu$m (and in the groundbased 
3\,$\mu$m spectrum; Imanishi \& Dudley \cite{Imanishi00}), 
while absorptions from water ice and silicates are found at 6.0\,$\mu$m 
and 9.7\,$\mu$m, respectively (Spoon et al. \cite{Spoon02}). The 
strongest feature in the spectrum, peaking at 7.7\,$\mu$m, has been 
previously attributed exclusively to PAH emission. On closer inspection,
however, the feature seems to be a blend of a 7.7\,$\mu$m PAH emission 
feature and a continuum peak resulting from absorption by ice on the 
short wavelength side and silicates on the long wavelength side. 

\begin{figure}[]
 \begin{center}
  \psfig{figure=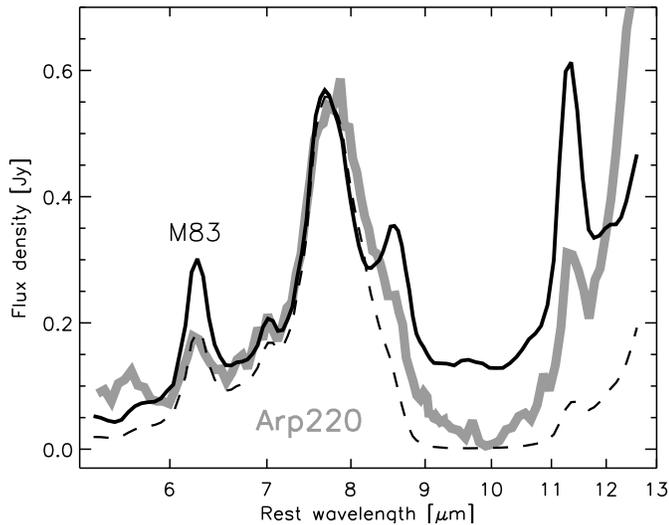,width=8.8cm,angle=0}
 \end{center}
\vspace*{-5mm}
\caption{Comparison of the spectra of Arp\,220 ({\it thick grey line})
and the central region of starburst galaxy M\,83 ({\it black line}).
The M\,83 spectrum shows strong PAH emission features at 6.2, 
7.7, 8.6 and 11.2\,$\mu$m. In contrast, in Arp\,220 the 6.2, 8.6 
and 11.2\,$\mu$m features are weak. The {\it black dashed line} 
illustrates the effect of strong extinction (A(V)=105; A(7.7\,$\mu$m)=1.2) 
on the M\,83 spectrum. Both versions of the M\,83 spectrum have been 
scaled to match the 7.7\,$\mu$m peak in the spectrum of Arp\,220.}
\label{arp220_m83}
\end{figure}

Close comparison of the 7.7\,$\mu$m feature in the spectrum of Arp\,220 
with the 7.7\,$\mu$m PAH feature in the spectrum of the starburst galaxy
M\,83 (Fig.\,\ref{arp220_m83}) shows that the 7.7\,$\mu$m feature in 
Arp\,220 is quite broad. Moreover, the ratio of the 7.7\,$\mu$m peak
to the 6.2\,$\mu$m and 11.2\,$\mu$m PAH bands is very large in Arp\,220 
compared to other sources (cf. Fig.\,\ref{arp220_m83}). The profile 
of 7.7\,$\mu$m PAH emission bands is, however, known to be very stable over 
a wide range of integrated galaxy spectra (Helou et al. \cite{Helou00}). 
Peeters et al. (\cite{Peeters02}) found that Galactic ISM spectra show 
a similarly stable 7.7\,$\mu$m PAH feature. Only for evolved stars and 
isolated Herbig AeBe stars are the width and central wavelength of the
7.7\,$\mu$m PAH feature known to vary (Peeters et al. \cite{Peeters02}). 
These stars are however very unlikely to dominate the spectrum
of Arp\,220. Therefore, there is no reason to assume that PAH 
features in Arp\,220, if present, should have an intrinsically different 
shape than in any other galaxy spectrum. Likewise, while the relative
strengths of the PAH features are known to vary from source to source,
no Galactic or extragalactic sources show such extreme ratios as Arp\,220
(Peeters et al. \cite{Peeters02}). Alternatively, the large width 
of the 7.7\,$\mu$m feature might also be the result of a strong velocity
dispersion within the PAH emitting gas ($\Delta$v$\sim$5000 km/s). 
A similar velocity broadening is, however, not observed for the 6.2 and 
11.2\,$\mu$m PAH features.
Strong extinction as a cause for the unusual strength and width of the 
7.7\,$\mu$m peak can also be ruled
out. This is best illustrated by fitting a starburst spectrum 
(here: M\,83) to the peak of the 7.7\,$\mu$m feature and applying
foreground extinction (Fig.\,\ref{arp220_m83}). 
The weakness of the 6.2\,$\mu$m PAH feature is reproduced well for a 
foreground extinction A(V)=105, but the spectrum beyond 8.5\,$\mu$m and
the 8.6\,$\mu$m and 11.3\,$\mu$m PAH features are not. While the relative
feature strengths depend on the adopted extinction curve (Sect.\,3.2),
no plausible extinction will be able to {\it widen} the 7.7\,$\mu$m feature.
We here propose instead the weakness of the 6.2\,$\mu$m PAH feature to 
be indicative of an unusually small contribution of the family of PAH 
emission features to the mid-IR spectrum of Arp\,220 and the unusual
strength and width of the 7.7\,$\mu$m feature to be the result of an 
underlying mid-IR continuum, peaking strongly near 7.7\,$\mu$m. 
Evidence in support for this model is presented below.

\subsection{Broad 7.7\,$\mu$m feature similar to Mon\,R2:IRS\,1+2}

\begin{figure}[]
 \begin{center}
  \psfig{figure=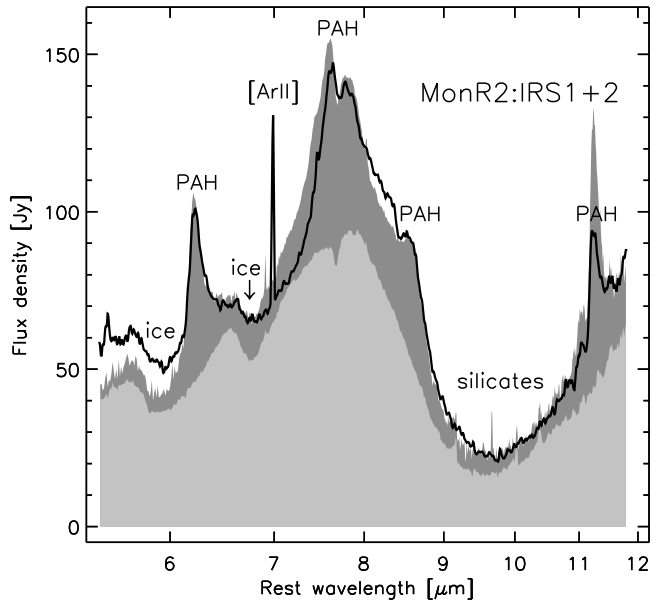,angle=0}
 \end{center}
\vspace*{-5mm}
\caption{The line of sight to the embedded protostar Mon\,R2:IRS\,1 
passes through the ultra compact $\ion{H}{ii}$ region Mon\,R2:IRS\,2.
The ISO--SWS spectrum of Mon\,R2:IRS\,1+2 ({\it black}) hence shows 
features of both type of environments: PAH emission from the ultra compact 
$\ion{H}{ii}$ region and a dust and ice absorbed continuum from the
embedded protostar. Here we show a simple 2-component fit to the 
Mon\,R2:IRS\,1+2 spectrum, using the ISO--SWS spectra of the 
protostar NGC\,7538:IRS\,9 ({\it light grey area}) and the reflection 
nebula NGC\,7023 ({\it dark grey area}) as fitting templates.}
\label{monr2_decomp}
\end{figure}

The shape of the 7.7\,$\mu$m feature in Arp\,220 appears to be unique 
among a sample of more than 250 galaxies observed spectroscopically
in the 6--12\,$\mu$m range (Spoon et al. in prep.). The feature has,
however, a Galactic counterpart: the ISO--SWS spectrum of the 
combined line of sight to the sources Mon\,R2:IRS\,1+2 
(Fig.\,\ref{monr2_decomp}). 
Mon\,R2 is a massive Galactic star formation region. Infrared
observations show several compact sources and extended emission in 
the central region of the giant molecular cloud (Beckwith et al. 
\cite{Beckwith76}). An elliptical ring encloses two IR sources, 
IRS\,1 and IRS\,2. IRS\,1, with a presumed spectral type of B0 
(Downes et al. \cite{Downes75}; Howard et al. \cite{Howard94}), 
is the exciting source of the compact $\ion{H}{ii}$ region enclosed 
by the IR ring (Massi, Felli, \& Simon \cite{Massi85}). IRS\,2 is 
still deeply embedded in the molecular cloud and most probably at 
an earlier stage of formation. Given this confused line of sight, 
the broad 7.7\,$\mu$m feature in the spectrum of Mon\,R2:IRS\,1+2 
may well be the result of the superposition of a strongly absorbed 
continuum, peaking at $\sim$8\,$\mu$m, and a `normal' 7.7\,$\mu$m 
PAH emission feature. Fig.\,\ref{monr2_decomp} 
shows the result of a crude decomposition of the Mon\,R2:IRS\,1+2 
spectrum into the ISO--SWS spectra of the embedded protostar 
NGC\,7538:IRS\,9 and the reflection nebula NGC\,7023. The fit is
quite good given the fact that the columns of ices and silicates
vary a lot from one embedded protostar to the other. Note that the 
8.6\,$\mu$m PAH feature is not suppressed by extinction, but instead
stands out as a shoulder on the flank of the 7.7\,$\mu$m feature,
filling in the blue wing of the 9.7\,$\mu$m silicate absorption 
feature.

\subsection{Mid-IR continuum similar to NGC\,4418}

\begin{figure*}[]
 \begin{center}
  \psfig{figure=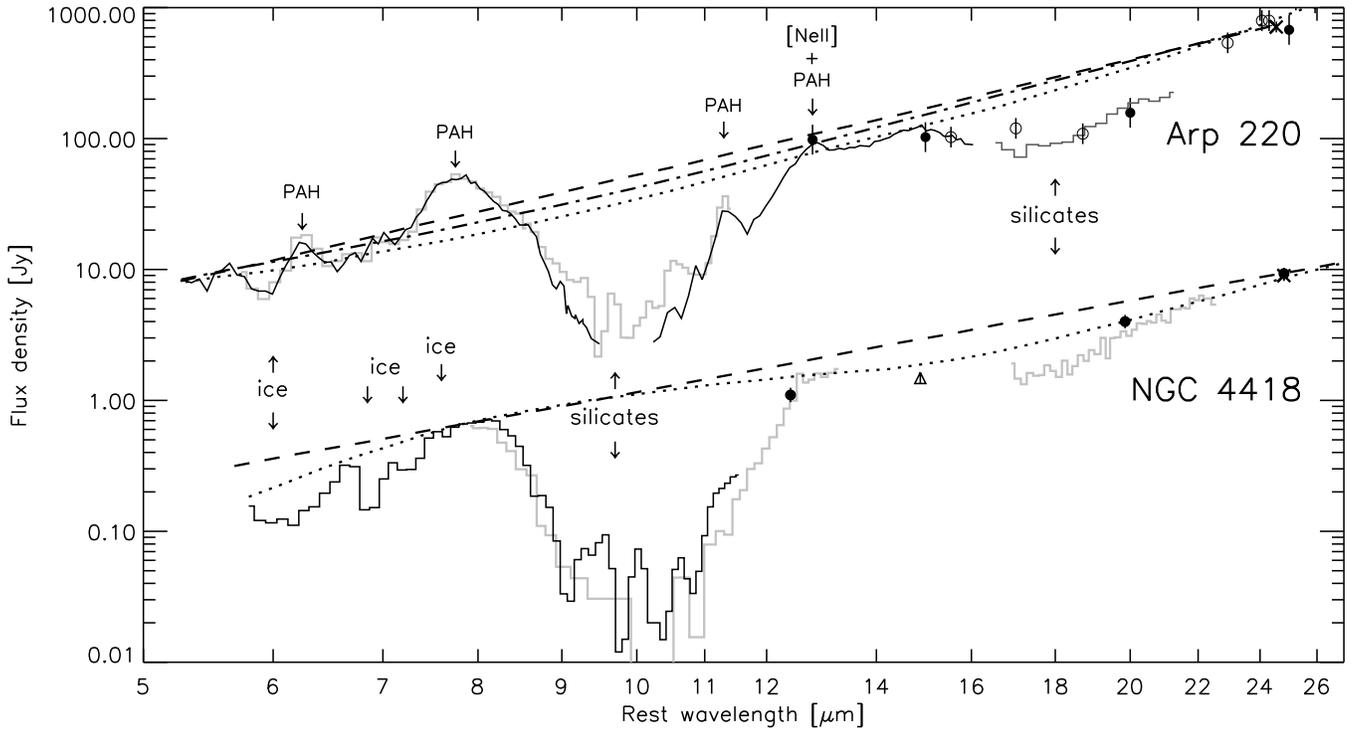,width=17.9cm,angle=0}
 \end{center}
\vspace*{-5mm}
\caption{Comparison of the mid-IR spectra of Arp\,220 (multiplied by 90) 
and NGC\,4418. For Arp\,220 the following spectra are plotted: in {\it black}
the 5--16\,$\mu$m ISO--CAM--CVF spectrum, in {\it light grey} the 
5.6--11.4\,$\mu$m ISO--PHT--S spectrum and in {\it dark grey} the 
17--22\,$\mu$m UCL spectrum (Smith et al. \cite{Smith89}). ISO--SWS 
continuum points are marked by {\it open circles}, ISO--PHT photometry by 
{\it filled circles} and the IRAS 25\,$\mu$m flux by a {\it cross}. For 
NGC\,4418 the spectra plotted are: in {\it black} the 5.6--11.4\,$\mu$m 
ISO--PHT--S spectrum and in {\it light grey} the 8--13\,$\mu$m 
and 17--23\,$\mu$m UCL spectra of Roche et al. (\cite{Roche86}). 
IRTF photometry (Wynn-Williams \& Becklin \cite{Wynn-Williams93}) is 
marked by {\it filled circles}, ISO--CAM (LW3) photometry by a 
{\it triangle} and the IRAS 25$\mu$m flux by a {\it cross}. The positions 
of absorption and emission bands are indicated, as are several choices 
for the local continuum for each object.}
\label{mirseds}
\end{figure*}

In Fig.\,\ref{mirseds} we compare the mid-IR spectra of Arp\,220 and NGC\,4418.
The spectrum of NGC\,4418 is dominated by strong silicate absorption bands
at 9.7\,$\mu$m and 18\,$\mu$m (Roche et al. \cite{Roche86}) and ice
absorption bands at 6.0\,$\mu$m (H$_2$O), 6.85\,$\mu$m \& 7.25\,$\mu$m 
(Hydrogenated Amorphous Carbons; HAC) and 7.67\,$\mu$m (CH$_4$)
(Spoon et al. \cite{Spoon01}). No mid-IR emission features, including the
commonly detected 12.81\,$\mu$m $[\ion{Ne}{ii}]$ fine structure line, 
have been detected so far (Spoon et al. \cite{Spoon01}; 
R.\,Siebenmorgen, unpublished TIMMI2 spectra). A first order estimate 
for the mid-IR local continuum of NGC\,4418 is obtained by fitting 
a power law through two feature-free ``pure continuum'' pivots at 
8.0\,$\mu$m and 25\,$\mu$m (the dashed continuum in Fig.\,\ref{mirseds}). 
Another, more conservative choice of continuum, assuming ``pure continuum'' 
at 6.7\,$\mu$m and 13\,$\mu$m too, is shown as a dotted line in 
Fig.\,\ref{mirseds}. 

The mid-IR spectrum of Arp\,220 (Fig.\,\ref{mirseds}) looks very similar 
to the pure absorption spectrum of NGC\,4418, except for the presence of 
weak emission features due to PAHs (6.2\,$\mu$m, 7.7\,$\mu$m and  
11.2\,$\mu$m), 6.91\,$\mu$m H$_2$ 0--0 S(5) and 12.81\,$\mu$m $[\ion{Ne}{ii}]$ 
(Sturm et al. \cite{Sturm96}; Genzel et al. \cite{Genzel98}). Using the 
same method as for NGC\,4418, we estimate the local continuum in Arp\,220 by 
fitting a power law through two feature-free ``pure continuum'' pivots
at 5.5\,$\mu$m and 25\,$\mu$m (the dashed continuum in Fig.\,\ref{mirseds}).
A more conservative estimate for the local continuum is obtained by 
including ``pure continuum'' pivots at 6.7\,$\mu$m and 14--15\,$\mu$m 
as well.
This results in the dotted continuum for Arp\,220. The dash-dotted 
continuum in Fig.\,\ref{mirseds} is a compromise between the two. 
Note that besides the obvious silicate absorption features at 9.7\,$\mu$m
and 18\,$\mu$m all three continua imply the presence of
an absorption feature due to water ice, which runs from 5.7\,$\mu$m
to 8.0\,$\mu$m (Spoon et al. \cite{Spoon02}). The emission features 
within this range, 6.2\,$\mu$m PAH, 6.91\,$\mu$m  H$_2$ 0--0 S(5),
7.7\,$\mu$m PAH and likely (though not targeted by ISO--SWS) 
7.0\,$\mu$m $[\ion{Ar}{ii}]$, fill up the absorption partially
or even turn it into emission.

Given the close similarity of the mid-IR continua of Arp\,220 and
NGC\,4418 and the presence of strong silicate and ice absorption 
features in both spectra, the mid-IR spectrum of Arp\,220 seems to be
the superposition of a strongly absorbed continuum and a typical 
PAH-dominated spectrum. In the next Section we will test this hypothesis 
by decomposing the mid-IR spectrum of Arp\,220 into a PAH-dominated 
spectrum and an absorbed continuum.

\section{Mid-IR spectral decomposition}

In order to test the superposition hypothesis, we have fit several 
combinations of a mid-IR absorbed continuum source and a PAH
template to the observed Arp\,220 mid-IR spectrum. Our method
differs from existing methods (e.g. Tran et al. \cite{Tran01}) 
by using observed templates instead of model mid-IR continua.
The latter usually do not take into account the complex radiative 
transfer effects due to ices and silicates that give rise to 
the exotic observed spectral shapes of the continuum sources 
and hence may fail to reproduce their spectra properly.

\subsection{PAH and continuum templates}

\begin{figure*}[]
 \begin{center}
  \psfig{figure=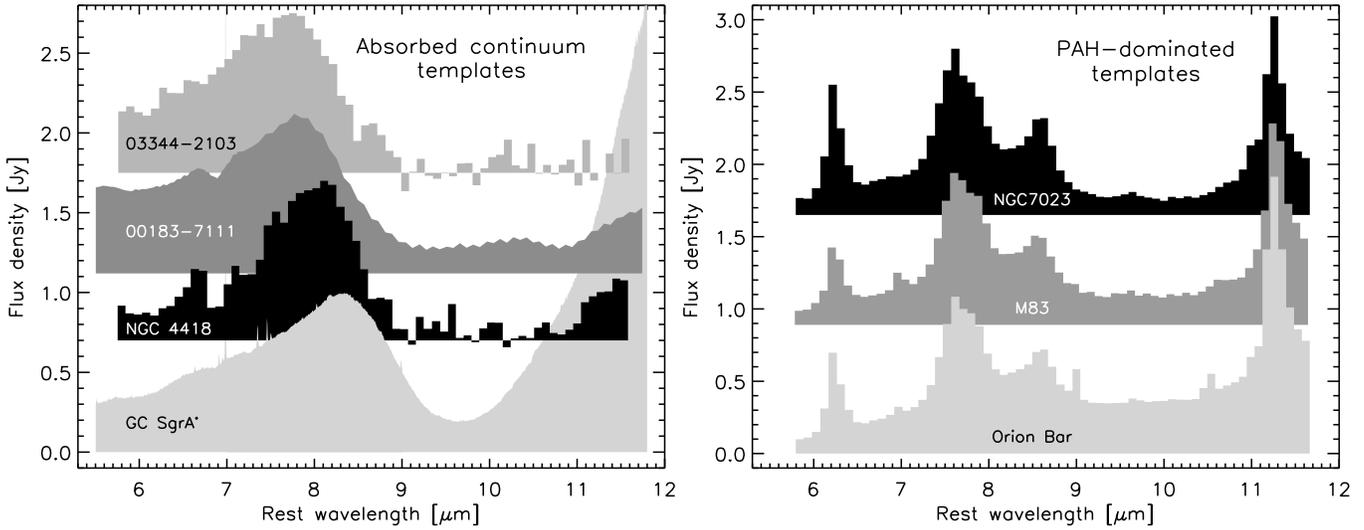,width=17.9cm,angle=0}
 \end{center}
\vspace*{-5mm}
\caption{PAH and continuum templates used in the decomposition
of the observed mid-IR spectrum of Arp\,220. {\bf Left panel:}
Four mid-IR absorbed continuum spectra. The spectra are
shown at their instrumental resolution: R$\sim$90 for 
IRAS\,03344--02103 and NGC\,4418; R$\sim$40 for IRAS\,00183--7111;
R$\sim$1000 for Sgr\,A$^*$. The spectra have 
been normalized to their 7.5--8.5\,$\mu$m peak fluxes and are sorted 
according to the central wavelength of their 7.5--8.5\,$\mu$m peak 
flux. {\bf Right panel:} Three mid-IR PAH-dominated spectra, shown 
at a spectral resolution R$\sim$90. The spectra have been normalized 
to their 7.7\,$\mu$m peak fluxes and are sorted in order of increasing 
9--12\,$\mu$m continuum.}
\label{templates}
\end{figure*}

The absorbed continuum sources span a range of spectral shapes 
(Fig.\,\ref{templates}a), with 6--12\,$\mu$m peak flux 
wavelengths ranging between 7.7\,$\mu$m for I\,03344--2103 and 
8.3\,$\mu$m for the Galactic Center (Sgr\,A$^*$). All four sources 
exhibit a strong 9.7\,$\mu$m silicate feature and three of them 
also show clear signs of ice absorption features. The silicate
optical depths range from $\tau_{\rm sil}\sim$1.9\footnote{ 
this is the aparent optical depth, measured from the ISO--SWS
spectrum (Fig.\,\ref{templates}a). Corrected for silicate
emission along the line of sight the true value is 3.6$\pm$0.3
(Roche \& Aitken \cite{Roche85}).}
for Sgr\,A$^*$ and $\tau_{\rm sil}>$1.9 for IRAS\,00183-7111 
(Spoon et al. \cite{Spoon02}) to $\tau_{\rm sil}>$3.5 for 
IRAS\,03344--2103 and $\tau_{\rm sil}>$3.7 for NGC\,4418. For the 
latter, Roche et al. (\cite{Roche86}) claim an even higher value, 
$\tau_{\rm sil}\sim$7. 
Judging from the steepness of the blue wing of its silicate feature
(Fig.\,\ref{templates}a), NGC\,4418 may well have the largest dust 
column of all four absorbed continuum templates.

As PAH templates we selected the reflection nebula NGC\,7023,
the Orion Bar star forming region and the central region of the
starburst galaxy M\,83. The latter has been preferred over other 
nearby starburst templates, like M\,82 or NGC\,253, because its 
5--16\,$\mu$m spectrum appears less affected by extinction 
than the other two galaxy spectra. Together, the three selected 
PAH templates (Fig.\,\ref{templates}b) are meant to cover 
the full range of spectral shapes from quiescent PDRs to intense
star forming regions. Note the large difference in strength of 
the 9--12\,$\mu$m continuum between the spectra of NGC\,7023 and 
the Orion Bar. The difference is attributed to the presence
of hot dust in the Orion $\ion{H}{ii}$ region, which is absent
in the cold environment of a reflection nebula. In contrast, the 
PAH spectra are quite similar, except for a somewhat weaker
6.2\,$\mu$m PAH feature relative to the 7.7\,$\mu$m PAH feature
in M\,83 and a somewhat stronger 11.2\,$\mu$m PAH feature 
relative to the 7.7\,$\mu$m PAH feature in the Orion Bar spectrum.

\subsection{The extinction law at mid-infrared wavelengths}

Unlike the extinction law at optical or near-IR wavelengths, 
surprisingly little is known about the shape and the general 
applicability of the extinction law at mid-IR wavelengths. 
Results obtained for different lines of sight vary considerably. 
A good definition of the shape of the mid-IR extinction law is, 
however, important for a good decomposition of the Arp\,220 spectrum.

\begin{figure}[]
 \begin{center}
  \psfig{figure=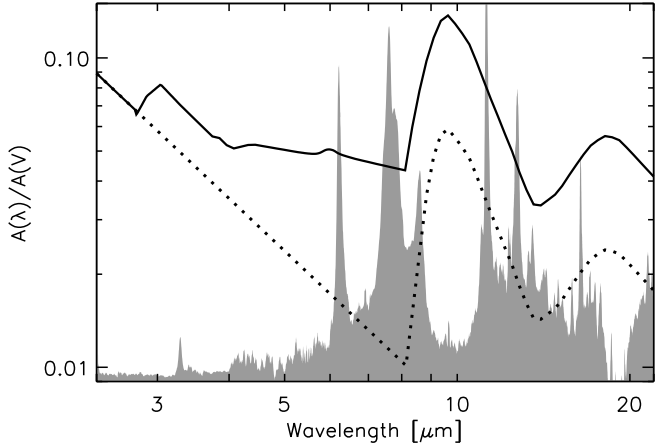,width=8.8cm,angle=0}
 \end{center}
\vspace*{-5mm}
\caption{Comparison of two mid-IR extinction laws, 
overlayed on the PAH emission spectrum of reflection nebula
NGC\,7023 ({\it grey area}).
The Galactic center extinction law of Lutz (\cite{Lutz99}) 
is shown as a {\it black line}, the extinction law of Draine 
(\cite{Draine89}) for the local solar neigbourhood as a 
{\it dotted line}. The extinction laws are refered to as 
{\tt Lutz\_gc} and {\tt draine\_local}, respectively.}
\label{extlaws}
\end{figure}

The differences among extinction laws appear largest at 3--8\,$\mu$m. 
Assuming a standard graphite-silicate mixture, a $\lambda^{-1.75}$ 
power law fall-off is expected for this wavelength range 
(Fig.\,\ref{extlaws}; Draine \cite{Draine89}; Martin \& Whittet 
\cite{Martin90}). ISO measurements of 
molecular hydrogen towards the Orion Peak-1 as well as $\ion{H}{i}$
recombination lines of compact $\ion{H}{ii}$ regions support 
this model (Bertoldi et al. \cite{Bertoldi99}; Mart\'{\i}n-Hern\'andez
et al. \cite{Martin03}). 
$\ion{H}{i}$ recombination line observations, probing the dusty, 
complex line of sight to the Galactic center (Sgr\,A$^*$), indicate, 
however, a nearly flat extinction curve between 3 and 8\,$\mu$m 
(Fig.\,\ref{extlaws}; Lutz \cite{Lutz99}), with
A$_{\lambda}$/A$_V$ a factor $\sim$4 higher at 7\,$\mu$m compared to 
the extinction law of Draine (\cite{Draine89}). This would point to
the presence of larger, fluffier grains in the line of sight to 
the Galactic center than towards other Galactic sources (e.g. Kr\"ugel
\& Siebenmorgen \cite{Kruegel94}).

Beyond 8\,$\mu$m, the 9.7 and 18\,$\mu$m bands of amorphous silicates 
dominate the extinction curve. Both the shape (FWHM) and the strength 
of the bands (both A$_{9.7}$/A$_V$ and A$_{9.7}$/A$_{18}$) are reported 
to vary between different lines of sight (e.g. Fig.\,\ref{templates}a) 
and from observer to observer (Draine \cite{Draine89}). Here we will
adopt the astronomical silicate profile of Weingartner \& Draine 
(\cite{Weingartner01}).

In order to explore the effect of differences between mid-IR extinction 
curves on the spectral decomposition, we here define two extinction 
curves which should be representative for the range of extinction 
properties between different lines of sight.
The extinction law, which we will refer to as {\tt Draine\_local},
combines the $\lambda^{-1.75}$ power law fall-off for $\lambda<$8.14\,$\mu$m
with a silicate feature of strength A$_{9.6}$/A$_V$=0.06 typical for the 
local solar neighbourhood (Roche \& Aitken \cite{Roche84}). 
In contrast, the {\tt Lutz\_gc} extinction law is `flat' from 
3--8\,$\mu$m and has a silicate feature strength of A$_{9.6}$/A$_V$=0.14
(Lutz \cite{Lutz99}). Both extinction laws are shown in Fig.\,\ref{extlaws}

\subsection{Decomposition method}

For each continuum and PAH template combination we have explored a four
parameter space for the best fit to the observed Arp\,220 ISO spectrum. 
The four parameters explored are:
\begin{itemize}
\item the contribution of the absorbed-continuum spectrum to the total 
      spectrum
\item the contribution of the PAH spectrum to the total spectrum
\item the adopted extinction law: either {\tt draine\_local} or
      {\tt Lutz\_gc}
\item the amount of reddening (A$_{\rm V}$) on the PAH spectrum
\end{itemize}

For a 2-component model spectrum to give a good fit to the Arp\,220 spectrum,
the model spectrum has to reproduce several key spectral features which make 
the Arp\,220 spectrum stand apart from other mid-IR galaxy spectra.
These key features are, in order of decreasing importance:
\begin{itemize}
\item the width of the 7.7\,$\mu$m feature
\item the absence of the 8.6\,$\mu$m PAH feature in the red flank of 
      the 7.7\,$\mu$m feature
\item the strength of the 11.2\,$\mu$m PAH feature relative to the 
      7.7\,$\mu$m feature
\item the strength of the 6.2\,$\mu$m PAH feature relative to the 
      7.7\,$\mu$m feature
\item the depth of the 9.7\,$\mu$m silicate feature
\item the profile of the 6.0\,$\mu$m water ice feature
\end{itemize}

\subsection{Decomposition results}

\begin{figure*}[]
 \begin{center}
  \psfig{figure=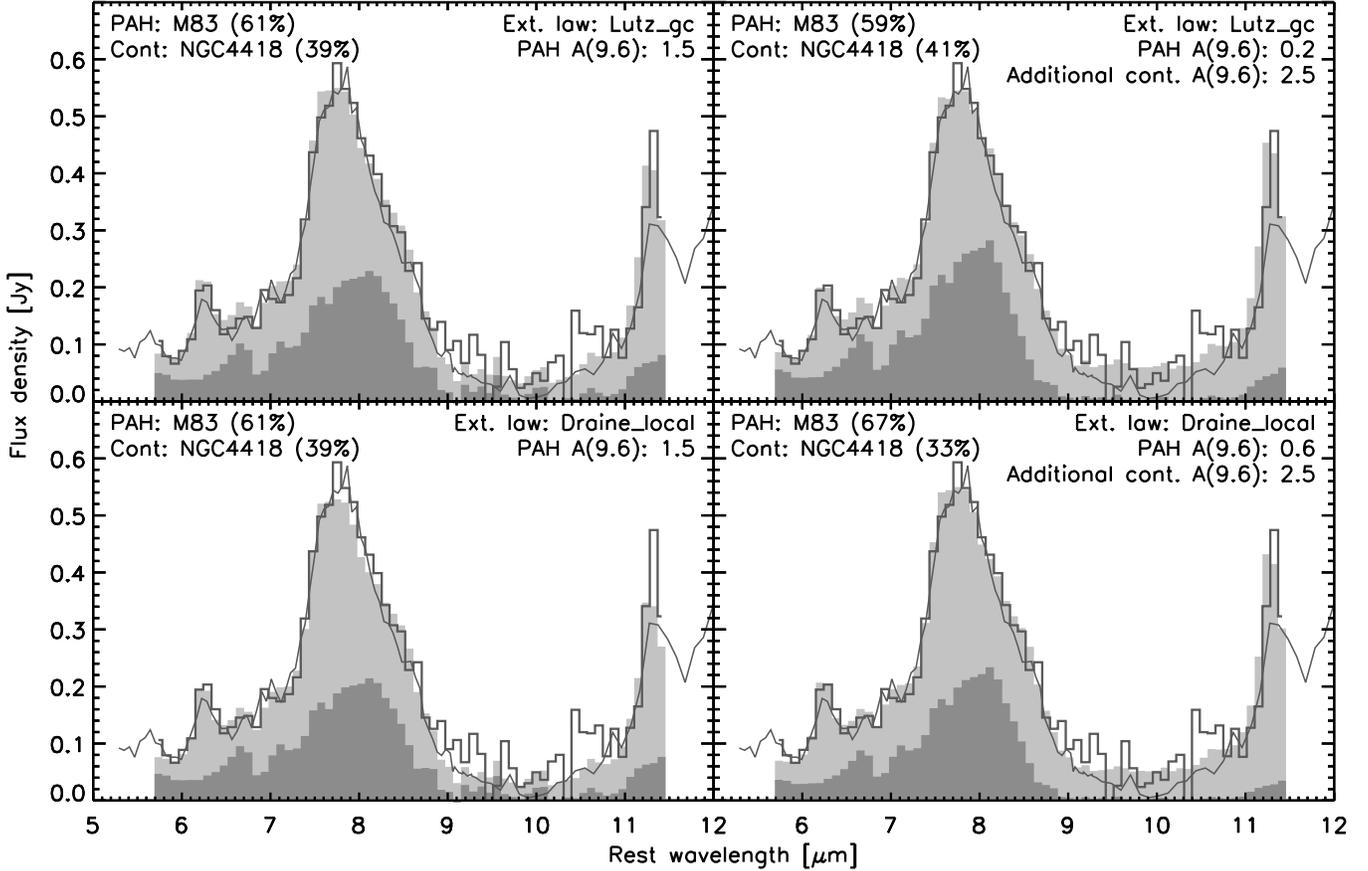,width=17.9cm,angle=0}
 \end{center}
\vspace*{-5mm}
\caption{Four 2-component fits to the Arp\,220 mid--IR ISO spectra 
(ISO--PHT--S: {\it dark grey histogram}; ISO--CAM--CVF: 
{\it dark grey line}).
In each panel the {\it dark grey area} represents the contribution of
the continuum component and the {\it light grey area} the contribution
of the PAH component. The PAH component is the same in all panels: the 
spectrum of the starburst galaxy M\,83. The absorbed continuum component 
differs between left and right panels. In the left panels it is NGC\,4418
as observed; in the right panels it is NGC\,4418 as seen through an 
additional screen of A(9.6\,$\mu$m)=2.5, equivalent to a factor ten
more attenuation at 9.6\,$\mu$m.
The fractional contribution of each fit component 
to the total 6--12\,$\mu$m fit is stated in brackets behind the name of 
the template. Also indicated is the amount of 9.6\,$\mu$m screen 
extinction on the PAH component.}
\label{best_fit_panel}
\end{figure*}

In Fig.\,\ref{best_fit_panel} we present the best fits to the 
6--12\,$\mu$m Arp\,220 spectrum. The fits clearly show that an appreciable 
contribution from a strongly absorbed continuum source (e.g. NGC\,4418,
or a stronger absorbed version of this spectrum) is required for a good 
fit to the observed spectral features. Compared to the extinction on 
the continuum source ($\tau_{\rm sil}>$3.7), the extinction on the PAH 
component is minor, ranging from $\tau_{\rm sil}$=0.2 to 1.4). At these 
small obscurations, the choice of extinction law does not play an 
important role.
The best fits further reveal that a direct measurement of the silicate 
optical depth from the ISO spectrum will be severely hampered by the 
presence of filled-in emission from the PAH component 
(see Fig.\,\ref{best_fit_panel}), resulting in a serious underestimation
of the true silicate optical depth.

Other combinations of our continuum and PAH templates fail to reproduce
key features of the Arp\,220 spectrum. Fits involving the continuum 
templates IRAS\,03344--2103 and IRAS\,00183--7111, for example, fail to 
fit the width of the 7.7\,$\mu$m feature, because their continua peak
at too short a wavelength (Fig.\,\ref{templates}a). The continuum of 
Sgr\,A$^*$, on the other hand, does peak at the right wavelength, but 
its silicate feature is too shallow and too narrow to leave much room 
for any contribution from the PAH component to the 8.5--11.5\,$\mu$m fit.
Strong extinction will have to be imposed on the PAH component to
minimize its contribution to the 8.5--11.5\,$\mu$m fit. This, however,
also minimizes the flux in the 11.2\,$\mu$m PAH feature to beyond what 
is still consistent with the observations. 
A similar problem exists for the Orion Bar PAH template. Its 
10\,$\mu$m continuum is the strongest among the three PAH
templates (Fig.\,\ref{templates}b) and hence requires an appreciable
extinction on its contribution to any Arp\,220 fit; too much for
a good fit to the 11.2\,$\mu$m PAH feature. PAH template NGC\,7023
suffers from the opposite problem. Its continuum is too weak to produce
good fits to the Arp\,220 spectrum. 

Absorptions by ices play an important role in distorting the spectral
shape of strongly dust enshrouded sources like NGC\,4418 
(Spoon et al. \cite{Spoon01}; Fig.\,\ref{templates}a). Their impact
is, however, small for moderately absorbed spectra like the line of 
sight to Sgr\,A$^*$ (Fig.\,\ref{templates}a). On the other hand, as 
the abundance of ices is variable and changes from one Galactic molecular 
cloud to another, it is useful to 
assess the impact of an increased water ice abundance on our fits. We 
therefore ran tests with a {\tt Lutz\_gc} extinction 
law with five times higher water ice abundance and compared the
best fitting parameters to those for the unmodified {\tt Lutz\_gc} 
extinction law. Only for those template combinations requiring high
foreground extinction on the PAH template were the individual parameters
found to change noticeably (but $<$10\%). For all other 
combinations, including our best fits (Fig.\,\ref{best_fit_panel}), 
the effects turn out to be negligible.
Further experiments with an extinction curve supplemented with both the
6\,$\mu$m ice and 6.8\,$\mu$m HAC absorption features show that under these 
conditions successful fits can be obtained also with an (additionally 
obscured) Sgr\,A$^*$ continuum. This stresses the presence of ice 
absorptions in the continuum as a key requirement for a successful fit, 
and one of the reasons why NGC\,4418 produces the best results in our 
original fits.

\section{Discussion}

Our successful decomposition of the 6--12\,$\mu$m ISO--PHT--S spectrum  
into a strongly absorbed continuum and a weakly absorbed PAH component 
confirms our initial suspicion that a) its exotic mid-IR spectrum 
resembles a blend of the spectra of NGC\,4418 and M\,83 
(Fig.\,\ref{photseds}) and b) that the only difference between the 
mid-IR spectra of Arp\,220 and a strongly ice- and dust-absorbed 
source like NGC\,4418 is the additional presence of PAH emission 
features in the spectrum of Arp\,220 (Fig.\,\ref{mirseds}).

\subsection{Identification of the spectral components}

The large difference in obscuration of the two spectral components of 
our fit indicates that these most likely represent two {\it spatially} 
separate components. Recent high angular resolution N-band spectroscopy 
of the nuclear region (Soifer et al. \cite{Soifer02}) show the 
11.2\,$\mu$m PAH emission
and the 11--12\,$\mu$m continuum emission to have clearly different
distributions. While the absorbed-continuum emission clearly peaks on 
the two nuclei, the PAH emission extends over a far wider area and does
not peak on either nucleus. We therefore associate our absorbed
continuum component with the two nuclei and our PAH component with
the region in between and around the two nuclei.

We estimate the infrared luminosity associated with the diffusely 
distributed PAH component from the observed 6.2\,$\mu$m PAH 
emission feature by assuming a generic conversion ratio between 
L(6.2\,$\mu$m PAH) and L(IR) and taking into account the weak
obscuration on the 6.2\,$\mu$m PAH feature as indicated by our 
best 2-component model fits. We derive the L(6.2\,$\mu$m PAH)/L(IR)
ratio from our sample of $\sim$70 mid-IR ISO spectra
of normal and starburst galaxies and obtain a value of 0.003$\pm$0.001.
Assuming this conversion factor applies also to the conditions in
the ULIRG Arp\,220 and applying it to our four best fitting template
combinations (Fig.\,\ref{best_fit_panel}) we find the infrared
luminosity associated with the diffuse PAH component to amount
to 1.2--2.1\,10$^{11}$\,L$_{\odot}$; 9--15\% of the total 
infrared luminosity of the system. 
A comparable result is obtained from the peak flux density of 
the 7.7\,$\mu$m PAH feature, using the empirical conversion factor
S(7.7\,$\mu$m PAH)/F(IR)=10$^{-11.84}$ Jy\,W$^{-1}$\,m$^2$ 
for starburst galaxies, determined by Lutz et al. (\cite{Lutz03}).
Taking into account the weak obscuration on the 7.7\,$\mu$m PAH 
feature (ranging from A(7.7\,$\mu$m)=0.07 to 0.49), the implied 
infrared luminosity is 0.7--1.8\,10$^{11}$\,L$_{\odot}$; 
5--13\% of the total infrared luminosity of the system. 
The results from both methods are in complete agreement with
the value derived by Soifer et al. (\cite{Soifer02}), who infered
a PAH-associated infrared luminosity of 1.2\,10$^{11}$\,L$_{\odot}$
(9\% of the total infrared luminosity) from their N-band spectrum.
Based on the small scatter between the three different methods,
we conclude that the infrared luminosity associated with 
the diffusely distributed PAH component in Arp\,220 amounts to 
5--15\% of the bolometric luminosity of the system, with some
uncertainty due to the trend towards a larger FIR/PAH emission
ratio with increasing interstellar radiation field intensity
in galaxies (e.g. Dale et al. \cite{Dale01}).

\begin{figure*}[]
 \begin{center}
  \psfig{figure=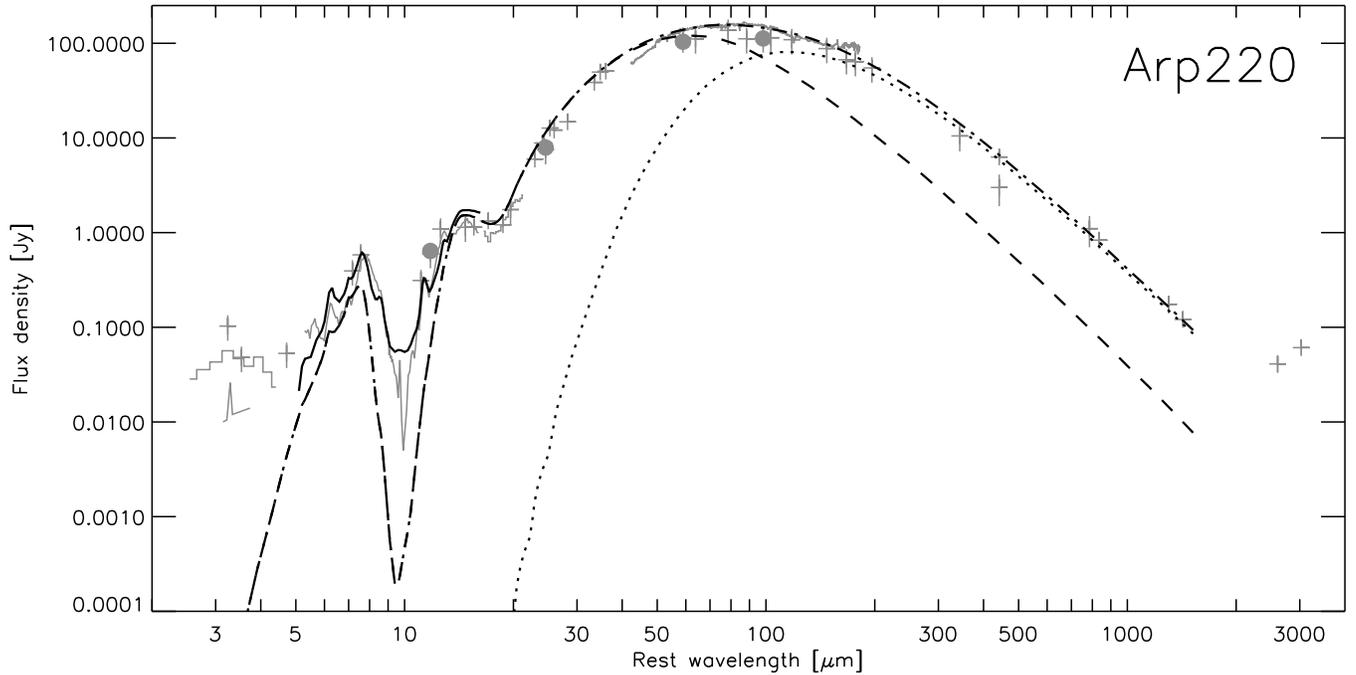,width=17.9cm,angle=0}
 \end{center}
\vspace*{-5mm}
\caption{The 2--3000\,$\mu$m spectrum of Arp\,220 ({\it grey})
overplotted with a three component fit. The {\it black dashed} curve 
represents the model SED for the two identical nuclei. The 
{\it black dotted} curve represents grey body emission from 30\,K cold 
dust. The {\it black dash-dotted} curve is the sum of these two components. 
The {\it black continuous} curve is the sum of the model SED and the 
weakly absorbed diffuse PAH component (M\,83) from our 2-component 
decomposition (Sect.\,3).}
\label{arp220_model_fit}
\end{figure*}

The literature reports the detection in the nuclear region
of several near- and mid-IR recombination and fine structure lines 
from star formation.
While no spatial information is available as to the origin of
the mid-IR lines (Sturm et al. \cite{Sturm96}), both the near-IR
recombination lines (Armus et al. \cite{Armus95}; 
Larkin et al. \cite{Larkin95}) and the radio recombination lines
(Anantharamaiah et al. \cite{Anantharamaiah00}) are mostly 
concentrated towards the two nuclei, rather than following the
diffuse PAH component. Extinction estimates range from A(V)$\sim$10 
in the near-IR (Armus et al. \cite{Armus95}; 
Larkin et al. \cite{Larkin95}) up to A(V)$\sim$40--60 in the 
mid-IR and radio (Sturm et al. \cite{Sturm96};
Anantharamaiah et al. \cite{Anantharamaiah00}), with noticable
uncertainties e.g. due to the measured line fluxes. This suggests
that the starburst activity traced by the emission lines includes
an intermediate obscuration component that is more concentrated
on the nuclei than the PAH emission. Two factors suggest this
component includes intermediate obscuration parts of the regions 
around the two nuclei rather than the deepest embedded parts:
A(V)$\preceq$50 appears still moderate compared to the extremely
obscured NGC\,4418-like continuum, and the energy budget of this
component stays a factor $\sim$2 below the bolometric luminosity
of Arp\,220 (Genzel et al. \cite{Genzel98}). The infrared 
emission lines may thus give a partial view of the circumnuclear
absorbed region, but very likely not a complete one.

\subsection{The nature of the nuclear power sources}

Given the relatively small fraction of the bolometric luminosity
associated with the diffuse PAH emission (5--15\%), the bulk of the 
luminosity of the Arp\,220 system must be associated with the 
absorbed-continuum component from the two nuclei. To be responsible 
for only $\sim$40\% of the 6--12\,$\mu$m luminosity but for 85--95\% 
of the total Arp\,220 infrared luminosity, these nuclei must, hence, 
be deeply enshrouded indeed. Given the large amounts of molecular 
material detected in the nuclear environment 
($\sim$10$^{10}$\,M$_{\odot}$ half of which is in the disk enclosing 
the two nuclei; Scoville et al. \cite{Scoville97}; 
Sakamoto et al. \cite{Sakamoto99}), this comes as no surprise.
Unfortunately, strong obscuration erases source-specific spectral 
signatures, not only at X-ray, UV, optical, near- and 
mid-IR wavelengths, but, depending on the absorbing column, 
also out to far-IR and, possibly, sub-mm wavelengths. 

Based on the few identified spectral signatures at hand, we consider 
two possibilities for the nature of the nuclear power sources.
First, each of the nuclei may contain a deeply embedded, extremely
dense and luminous stellar cluster -- a super-star cluster containing 
some 10$^6$ massive O stars within a region less than 100\,pc in size. 
Less extreme deeply embedded clusters have been discovered in the
starburst galaxies NGC\,5253 and He\,2--10 
(Gorjian et al. \cite{Gorjian01}; Vacca et al. \cite{Vacca02}).
A stellar nature for the nuclear power sources 
in Arp\,220 is further supported by the 18\,cm VLBI observations of 
Smith et al. (\cite{Smith98},\,\cite{Smith99}), which show a dozen
or so sources scattered over the two nuclei. 
These ``knots'' are consistent with free-free emission from luminous 
radio supernovae expanding in the dense (circum)nuclear environment 
(Smith et al. \cite{Smith98},\,\cite{Smith99}). 
Second, we consider two deeply embedded AGNs. Despite the lack of AGN 
features in any waveband, a deeply embedded powerful AGN at the center 
of each nucleus cannot be ruled out. The column density required to block 
the AGNs from detection by {\it BeppoSAX} in hard X-rays is 
N$_{\rm H_2}$=10$^{25.1}$--10$^{25.3}$\,cm$^{-2}$ with a large covering
factor (Iwasawa et al. 
\cite{Iwasawa01}). The non-detection of a high-T$_b$ radio core in
the 18\,cm VLBI maps of Smith et al. (\cite{Smith98},\,\cite{Smith99}) 
would then imply the AGNs to be radio-quiet or strongly free-free
absorbed.

\subsection{Modeling of the nuclear continuum}

In order to test whether the low mid-to-far-IR continuum ratio in Arp\,220 
may be attributed to strong dust obscuration on the emission of two identical,
deeply buried, energetically dominant, nuclear sources, we used the dust 
radiative transfer code of Siebenmorgen et al. (\cite{Siebenmorgen99},
\cite{Siebenmorgen01}) to model the 2--3000\,$\mu$m nuclear spectrum.

In the model we assume a cluster of OB stars, with stellar densities 
similar to those inferred for ultra-dense $\ion{H}{ii}$ regions 
(UD $\ion{H}{ii}$ regions; Vacca et al. \cite{Vacca02}), to reside at the 
center of each nucleus. For a given luminosity of 6\,10$^{11}$\,L$_{\odot}$,
the stars will occupy a sphere with radius r=25\,pc.
Adopting a nuclear gas mass of 10$^9$\,M$_{\odot}$, spherical symmetry,
constant gas density and a silicate optical depth of $\tau_{\rm sil}$=4.6 
(consistent with the silicate optical depth for our best fitting absorbed 
continuum source NGC\,4418), we are forced to place the outer radius of
the dust shell as far out as $\sim$325\,pc in order to lower the dust 
temperature to values consistent with the observed 12--50\,$\mu$m SED. 
The FWHM of the 11\,$\mu$m light profile is $\sim$20\,pc, consistent with 
the observations of Soifer et al. (\cite{Soifer02}).
As the temperature in the dust shell does not drop below 40\,K at the 
outer edge, we consider an additional cold dust component to account for 
the coldest dust in the system. This component likely accounts for 
the far-IR/submm emission associated with the diffuse PAH component and 
for far-IR/submm emission from the nuclear region not covered by our 
`simple' model. Here, we characterize the cold dust component by a grey body 
spectrum of 30\,K with a dust emissivity index of 1.8 and a dust optical 
depth of 1 at 100\,$\mu$m. Both the model spectrum of the nuclei and the 
grey body spectrum are shown in Fig.\,\ref{arp220_model_fit}. 
The Figure also shows the assumed contribution of the diffuse PAH 
component (M\,83; see Sect.\,3.4) to the Arp\,220 mid-IR spectrum. 

We conclude from our modeling that the low ratios 
S(6\,$\mu$m)/S(60\,$\mu$m) and S(6\,$\mu$m)/S(100\,$\mu$m) can, {\it in 
principle}, be explained by the effects of strong dust obscuration on
two deeply buried, energetically dominant, nuclear sources.
However, this model predicts a hydrogen column of only 
$\sim$10$^{23.2}$\,cm$^{-2}$
and hence favours stellar heating, as this column is a factor $\sim$100 too 
small to account for the upper limits on the hard X-ray flux if the total 
luminosity were dominated by AGN activity (Iwasawa et al. \cite{Iwasawa01}).

Note that if any deeply hidden AGN (N$_{\rm H}\geq$10$^{25}$ cm$^{-2}$) 
were present, the covering factor of the obscuration would have to be 
large in order to be consistent with the lack of reflected X-ray light 
measured by {\it Beppo}SAX (Iwasawa et al. \cite{Iwasawa01}). The AGN would 
hence strongly contribute to the far-IR continuum, but could not be 
responsible for the mid-IR continuum.

\subsection{Mid-to-far-infrared spectral characteristics}

Arp\,220 is an outlyer in many spectroscopic diagnostic diagrams linking 
mid-IR to far-IR quantities. In all cases Arp\,220 stands out
by having a low ratio of the mid-IR characteristic with respect to the 
far-IR characteristic:
\begin{itemize}
\item Already in the Bright Galaxy Sample of Sanders et al. 
(\cite{Sanders88}), Arp\,220 stands out by the smallest S12/S60 ratio. 
\item Similarly, Arp\,220 has the lowest S5.9/S60 ratio among the
larger sample studied spectroscopically in the mid-IR by Lutz
et al. (\cite{Lutz98}). \item The ratio of F(6.2\,$\mu$m PAH)/F(FIR) is 
lower than for any other 
galaxy in our database of more than 250 ISO mid-IR galaxy spectra. 
\item The ratio F($[$\ion{Ne}{ii}$]$)/F(FIR) is a factor three lower than
for the average starburst galaxy (Genzel et al. \cite{Genzel98}).
\item The ratio F(7.7\,$\mu$m PAH)/F(850\,$\mu$m) is the lowest among
a sample of normal and ultra-luminous galaxies (Haas et al. \cite{Haas01}).
\end{itemize}
Our spectral decomposition offers a simple explanation for the above 
observations. A typical starburst spectrum associated with the extended
PAH component will contribute strongly to the mid-IR but weakly to the
far-IR, while the dominant deeply enshrouded nuclei contribute weakly
to the mid-IR and strongly to the far-IR. The resulting combined
3--1000\,$\mu$m SED is hence mostly starburst-like in the mid-IR and 
dominated 
by cold dust emission from the enshrouded nuclei in the far-IR. Ratios 
of mid-IR to far-IR quantities, like F($[$\ion{Ne}{ii}$]$)/F(FIR),
F(6.2\,$\mu$m PAH)/F(FIR) or F(7.7\,$\mu$m PAH)/F(850\,$\mu$m), will hence 
all be systematically lower than the values typically found for starburst 
galaxies.
Interestingly, Luhman et al. (\cite{Luhman03}) advocate a similar scenario
as a contributor to the low F($[$\ion{C}{ii}$]$)/F(FIR) ratio of Arp\,220.

Other galaxies may exist with even more strongly obscured nuclei, emitting 
an even smaller fraction of the nuclear luminosity in the mid-IR than 
Arp\,220 does. These galaxies would hence look starburst-like in 
the mid-IR (contributed by a $[$weakly obscured$]$ circumnuclear starburst)
but would be characterized by a very strong cold dust continuum in the far-IR.
Galaxies of this type, may be recognized spectroscopically
by their low S5.9/S60, S5.9/S100, S5.9/S850 or F($[$\ion{C}{ii}$]$)/F(FIR)
ratios and their starburst-like 6.2\,$\mu$m-PAH line-to-continuum ratios. 
Perusal of our ISO spectral 
database has led to the identification of several candidate galaxies.
Arp\,220 may thus be a local and less extreme template of the class of 
SCUBA sources seen in the Hubble Deep Field.\\

Arp\,220 is not the only galaxy with spectral structure in the 6--12\,$\mu$m 
range reminiscent of strongly modified PAH bands. Other examples are 
Mrk\,231 and most of the galaxies in the sample of Tran et al. (\cite{Tran01}).
Some of these spectra show, in addition
to strong silicate absorption longward of 7.7\,$\mu$m, clear signs of 
water ice absorption shortward of 7.7\,$\mu$m. Depending on the 
strength of the PAH emission features these galaxies have been 
classified as class I, II or III ice galaxies (Spoon et al. \cite{Spoon02}).
Like Arp\,220, their spectra may well be the result of strong extinction 
on one or more spectral components.

\section{Conclusions}

We have shown that the 6--12\,$\mu$m spectrum of Arp220 is not that of 
a scaled-up typical starburst galaxy but contains a 'normal' 
starburst component characterized by PAH emission features plus a 
highly obscured dust continuum with ice and silicate absorption. 
Attempts to decompose the spectrum using a variety of extragalactic 
and Galactic template spectra yields a best fit in which a typical 
starburst, represented by M\,83, contributes $\sim$60\% and and an ice 
absorbed continuum galaxy, represented by NGC\,4418, $\sim$40\% of the 
6--12\,$\mu$m luminosity.  An important result in relation to previous 
studies is our conclusion that the pronounced emission feature 
peaking around 7.7\,$\mu$m is a blend of PAH emission and a broader 
feature in the continuum caused by ice absorption at shorter and 
silicate absorption at longer wavelengths.  We tentatively conclude 
that the PAH emitting component is only weakly absorbed and arises in 
the extended region imaged at higher resolution around 10\,$\mu$m by 
Soifer et al. (\cite{Soifer02}) whereas the absorbed continuum is 
associated with one or both of the compact nuclei. This extended starburst 
component contributes only 5--15\% of the total luminosity with the 
bulk emitted by the heavily obscured nuclear component(s).  One 
possibility is that this luminosity is generated by starburst 
activity occuring in a higher density environment than found in lower 
luminosity starburst galaxies due to the larger quantity of molecular 
gas and dust funnelled to the center by merging of the two nuclei. 
Due to the high extinction, it is 
not possible with the available data to exclude that AGN(s) also 
contribute some or all of this luminosity. Based on the upper limits 
for hard X-ray emission (Iwasawa et al. \cite{Iwasawa01}), however, 
Arp\,220 would need to be the most highly obscured AGN known.

\acknowledgements The authors wish to thank Tom Soifer and Eiichi Egami 
for sharing data with us and George Helou, Olivier Laurent, Matt Lehnert, 
Neil Nagar, Dave Sanders, Eckhard Sturm and Jacqueline van Gorkum for 
discussions. We are grateful to the referee for valuable suggestions.
This research has made use of the NASA/IPAC Extragalactic Database, which 
is operated by the Jet Propulsion Laboratory, Caltech under contract with 
NASA.

\end{document}